# Random sequential adsorption: from continuum to lattice and pre-patterned substrates

A Cadilhe[1], N A M Araújo[1] and Vladimir Privman[2]

[1] GCEP-Centro de Física da Universidade do Minho, 4710-057 Braga, Portugal
[2] Center for Advanced Materials Processing, Clarkson University, Potsdam, NY 13699-5721, USA

**Abstract**

The random sequential adsorption (RSA) model has served as a paradigm for diverse phenomena in physical chemistry, as well as in other areas such as biology, ecology, and sociology. In the present work, we survey aspects of the RSA model with emphasis on the approach to and properties of jammed states obtained for large times in continuum deposition versus that on lattice substrates, and on pre-patterned surfaces. The latter model has been of recent interest in the context of efforts to use pre-patterning as a tool to improve self-assembly in micro- and nanoscale surface structure engineering.

## 1. Introduction

Several recent experimental efforts [1–16] have focused on approaches of pre-treating two-dimensional (2D) surfaces, or one-dimensional (1D) structures, e.g., polymers, by imprinting microscale, and, ultimately, nanoscale, patterns, for producing new functional substrates for self-assembly and other applications. Micron and submicron colloidal particles have traditionally proven to be quite versatile building blocks for the construction of advanced materials. Thin films of adsorbed colloidal particles are of great technological interest for a wide range of devices, such as photonic crystals [17–19], quantum dots [20, 21], and heterogeneous catalysts [22, 23]. Kinetics of the deposition of colloidal and nanoparticles at surfaces also poses interesting problems from the theoretical point of view [24–28], as most of the above-mentioned applications require control of positioning and distances between neighbouring particles. Here we survey the model of random sequential adsorption (RSA), reviewed in [29, 30, 28], and outline some new results for deposition on pre-treated substrates. In the present work, we emphasize the use of the RSA model for irreversible monolayer surface deposition of well defined particles. However, the model has also been extended and applied in other contexts, in biology, ecology, sociology, and condensed matter physics [24, 26, 31, 27, 28, 32].

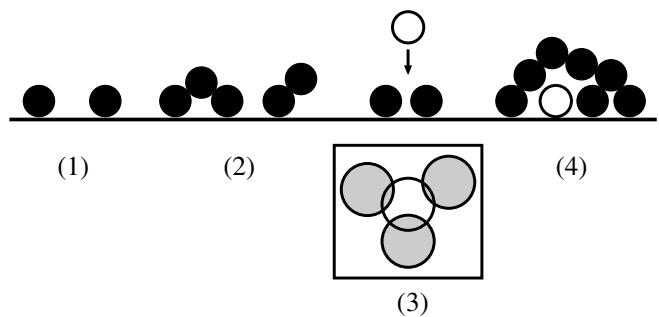

**Figure 1.** Possible configurations for particle deposition on a surface (that need not be planar). (1) Particles deposited onto the collector at low densities. (2) 'Multilayer' deposition on top of earlier deposited particles. (3) Jamming: deposition attempts of particles that do not fit in available gaps (voids) are rejected. The inset shows a 2D configuration. (4) Screening of part of the collector surface by earlier deposited particles: the position marked by an open circle is not reachable.

As already pointed out, surface deposition of submicron particles is of immense practical importance; see, e.g., [29, 30, 28, 33]. Particles of this size, colloid, protein, or even smaller nanoparticles and molecules, are suspended in solution, without appreciable sedimentation due to gravity. In order to maintain the suspension stable, one has to prevent aggregation (coagulation), that results in larger flocks for which the gravity pull is more profound. Stabilization by particle–particle electrostatic repulsion or by steric effects, etc, is usually effective for a sufficiently dilute suspension. Particles can then be deposited from solution, by diffusion, or by convective diffusion [34] from a flowing suspension, on collector surfaces. The suspension itself need not be dense even though the on-surface deposit might be quite dense, depending on the particle–particle and particle–surface interactions.

Figure 1 illustrates possible configurations of particles at a surface. From left to right, we show particles deposited on the surface of a collector, then particles deposited on top of other particles. The latter is possible only in the absence of particle–particle repulsion. The two situations are termed monolayer and multilayer deposition, even though the notion of a layer beyond that exactly at the surface is only approximate. We next show two effects that play an important role in surface growth. The first is jamming: a particle marked by an open circle cannot fit in the lowest layer at the surface. A more realistic two-dimensional (2D) configuration is shown in the inset. The second effect is screening: the surface position marked by the open circle is not reachable. Typically, in colloid deposition monolayer or few-layer deposits are formed and the dominant effect is jamming. Screening plays a dominant role in deposition of multiple layers and, together with the transport mechanism, determines the morphology of the growing surface. In addition, the particle configuration at the surface depends on the transport mechanism of the particles to it and on the particle motion on the surface, as well as possible detachment. Particle motion is typically negligible for colloidal particles but may be significant for proteins.

An important new development in surface deposition has been formation of controlled structures: the process of self-assembly. Here we will focus only on monolayer deposition. Obviously, the simplest way to have a structured deposit is to pre-pattern the surface in such a way that (the centre of) each arriving depositing object, particle or molecule, will fit at a single 'cell' or attachment site. These attachment sites can form a regular pattern and impose the same ordering on the depositing layer. However, with nano-sizes involved, such a precise patterning is not always feasible. More realistically, the substrate can be at best prepared in a state somewhere in between the uniform surface, the fully patterned one, and the disordered one: The imprinted (for instance, lithographically) features may be larger than the depositing objects, especially if the latter are molecules, and there may be some randomness in the surface 'cell' pattern. The features can also be smaller than the depositing particles, especially for colloid deposition. In the latter case, deposition of objects exactly fitting several lattice sites of a regular lattice structure has been extensively studied in the RSA literature [29, 30, 28, 33]. Recently, we have initiated studies that aim at exploring the effects of jamming and understanding the attachment kinetics of particles on pre-patterned substrates [35, 36].

In section 2, we introduce the continuum RSA model. Lattice RSA is then addressed in section 3. Section 4 introduces new features of deposition on a pre-patterned substrate with focus on the case on an exactly solvable Bethe-lattice model. Finally, section 5 is devoted to numerical results for correlations in and the nature of the jammed state for 2D patterned substrates, and to concluding remarks.

## 2. Continuum RSA, the role of the dimensionality and particle shape

The random sequential adsorption (RSA) model, see, e.g., [37, 24], assumes that particle transport (incoming flux) to the surface results in a uniform deposition attempt rate per unit time and area, $R$. In the simplest formulation, one assumes that only monolayer deposition is allowed. Within this monolayer deposit, each new arriving particle must either fit in an empty area allowed by the hard-core exclusion interaction with the particles deposited earlier, or its deposition attempt is rejected. The simplest case is that of continuum (off-lattice) deposition of spherical particles. However, other RSA models have also received attention. In 2D, noncircular cross-section shapes as well as various lattice-deposition models were considered [37, 24]. Several experiments on polymers and attachment of fluorescent units on DNA molecules [38] suggest consideration of the lattice–substrate RSA processes in 1D. RSA processes have also found applications in traffic problems and certain other fields. Our presentation in this section aims at off-lattice RSA models and outlines characteristic features of their dynamics.

Thus, we consider the fully irreversible RSA without detachment or diffusion. The substrate is usually assumed to be initially empty, at $t = 0$. For later times, $t > 0$, the coverage, $\theta(t)$, defined, e.g., by the fraction of the area covered by particles, increases and builds up to a fraction of unity on timescales of order $(RA)^{-1}$, where $A$ is the particle $D$-dimensional cross-sectional area. For deposition of spheres of radius $r$ on a planar (2D) surface, $A$ is actually the cross-sectional area $\pi r^2$, whereas the density of the deposited particles per unit surface area is given by $\theta(t)/A$.

At large times, the coverage approaches the jammed-state value, where only gaps too small to fit new particles were left in the monolayer. The resulting state with $\theta(\infty) < 1$ is less dense than the fully ordered close-packed coverage. The time-dependence of the RSA coverage is illustrated schematically in figure 2. At early times the monolayer deposit is not dense and the deposition events are largely uncorrelated. In this regime, mean-field low-density approximation schemes are useful [39–41].

The most interesting properties of the RSA process are the time dependence of the approach to the jammed state at large times, and the density correlations in the random, 'jammed-state' deposit that results. Deposition of $k$-mer particles on the linear lattice in 1D was in fact solved exactly for all times [42], and this solution yields the continuum 1D deposition as $k \to \infty$. In 2D, extensive numerical studies were reported [41, 43–54] of the variation of coverage with time and large-time asymptotic behaviour for both lattice and off-lattice models. Some exact results [42] for correlation properties are available in 1D. Numerical results for correlation properties, see, e.g., [45], have been obtained in 2D.

For continuum off-lattice deposition, the approach to the jamming coverage is power-law. This interesting behaviour [55, 56] is due to the fact that for large times the remaining gaps (voids) accessible to particle deposition can be of sizes arbitrarily close to those of the depositing particles. These gaps are then reached with very low probability by the depositing particles. The resulting power-law behaviour depends on the dimensionality and particle shape. For instance, for $D$-dimensional cubes of volume (area in 2D) $A$,

$$\theta(\infty) - \theta(t) \sim \frac{[\ln(RAt)]^{D-1}}{RAt}, \tag{1}$$

whereas for spherical particles,

$$\theta(\infty) - \theta(t) \sim (RAt)^{-1/D}, \tag{2}$$

where, as mentioned earlier, $A = \pi r^2$ in 2D.

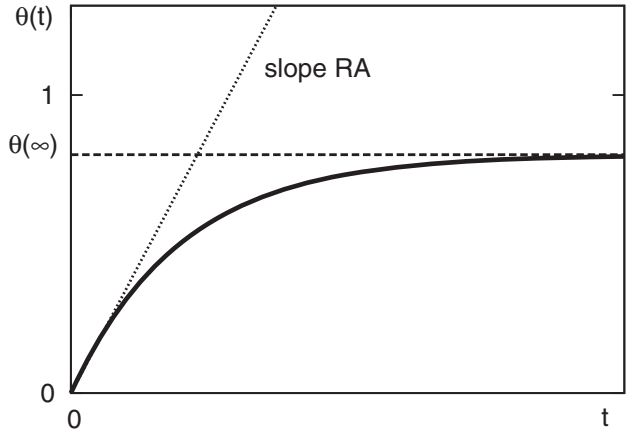

**Figure 2.** Schematic illustration of the time-dependence of RSA coverage.

The 1D asymptotic $(RAt)^{-1}$ law for deposition of segments of length $A$ (the car-parking problem) is confirmed by the exact solution [42]. For $D > 1$, such asymptotic laws were derived [54–57] based on the late-stage gap-distribution considerations. We do not review the details, but only point out that these empirical asymptotic relations obviously depend on the particle shape, rotational freedom upon deposition, and dimensionality of the substrate [54–57]. They have been verified, mostly in 2D, by extensive numerical simulations [28, 43–54].

The most studied 2D geometries are circles (corresponding to the deposition of spheres on a planar substrate) and squares. The jamming coverages are

$$\theta_{\text{circles}}(\infty) \simeq 0.544\text{–}0.550, \tag{3}$$

and

$$\theta_{\text{squares}}(\infty) \simeq 0.5620, \tag{4}$$

which are much lower that the respective close-packing values, $\pi/(2\sqrt{3}) \simeq 0.907$ and 1.

The correlations in the large-time jammed state are different from those of the equilibrium random gas of particles with density near $\theta(\infty)$. In fact, the two-particle correlations in continuum deposition develop a weak singularity at contact, and correlations generally reflect the full irreversibility of the RSA process [42, 45, 56].

## 3. Lattice RSA

Continuum deposition can be viewed as a limit of lattice RSA. For example, consider deposition of hypercubes on a hypercubic lattice with unit cell of size $L^D$. Let us assume that $A^{1/D}/L = k$, i.e., the deposited objects are $k \times k \times \cdots = k^D$ lattice cells in size. In the limit $k \to \infty$, $L \to 0$ with the product $kL$ fixed (thus, $A$ fixed), off-lattice deposition of cubic objects is recovered [42, 54]. This crossover has been investigated in some detail analytically, for the large-time asymptotic behaviour for any $D$, as well as numerically in 2D [54].

The new features of the lattice models are as follows. The gaps (voids) for single particles to land, which dominate the large-time asymptotics, are restricted by the lattice and can no longer be only slightly larger than the particles: the gap size increases by lattice increments from the exact fit on. As a result [54], the asymptotic convergence to the jamming coverage is exponential. Furthermore, it has been argued [54] that the relation replacing equation (1) takes the form

$$\theta(\infty; k) - \theta(t; k) \sim \Omega L^D e^{-RL^D t}, \qquad (5)$$

where the coefficient $\Omega$ has units of density.

It is important to define some conventions used in writing relations like equation (5). We note that, in order to keep the particle flux to the surface constant for varying $k$, we have to assume that particle deposition attempt events are 'registered' with the lattice, i.e. that the arriving particles 'wiggle in' to attempt deposition exactly at the discrete lattice locations with the ($k$-dependent) rate (per unit time) $RL^D = RA/k^D$. Furthermore, as already mentioned, $L^D = A/k^D$ is also a $k$-dependent quantity. The exponent in (5) is therefore given by $(RAt)/k^D$, and this exponential behaviour sets in for times of order $k^D(RA)^{-1}$. However, the coefficient $\Omega$ has no significant $k$ dependence [54].

Except for solvable 1D cases, these expectations have only been tested numerically (in 2D) for the case of RSA of lattice squares, and a semi-quantitative level of verification can be claimed [54]. However, these considerations do suggest that when the lattice mesh is of size $L^D$, much smaller than the sizes of the deposited objects, $A$, continuum-like behaviour will be approximately observed initially and up to rather large times on the scale of $1/(RA)$, for which the off-lattice power law will be manifest. However, ultimately, for times of order $k^D/(RA)$ and larger, exponential convergence of the type (5) will develop.

It is important to emphasize that the jamming coverage itself, $\theta(\infty; k)$, is dependent on the lattice and particle geometry and sizes, here collectively denoted by the added $k$-dependence. Relation (5) is for fixed geometry (fixed $k$) and increasing times. However, one can also explore the $k$-dependence of $\theta(\infty; k)$. Obviously, when the lattice cells and particles exactly match, there will be no jamming effect and we will have $\theta(\infty; 1) = 1$. A natural expectation is that the jamming coverage will then decrease monotonically with the increasing ratio of the linear particle to lattice-cell dimensions, $k$ in our example. Indeed, exact 1D results [42] and numerical estimates in 2D [54] confirm these expectations and suggest that

$$\theta(\infty; k) - \theta(\infty; \infty) \sim k^{-1}. \qquad (6)$$

There are currently no detailed studies of particle–particle correlations in the jammed state for varying $k$, to explore quantitatively to what extent a deposit with a coarser lattice mesh (smaller $k$) can be viewed as more 'ordered' than one with large $k$. However, if the main goal is to improve the resulting deposit density, then use of $k$ values of order unity is required. For example, for 2D squares we have the jamming coverage differences $\theta(\infty; k) - \theta(\infty; k+1) \sim 0.252, 0.068, 0.032$, for $k = 1, 2, 3$, respectively, which illustrates the dramatic effect of lattice confinement when the surface mesh limits the geometry of possible particle landing slots to 'cells' closely matching particle dimensions.

## 4. RSA on a pre-patterned Bethe lattice

A lattice can be pre-treated with patterns less regular than a lattice structure. One approach is to randomly block lattice sites or groups of sites. Correlations in such blockings can be defined, e.g., by considering pre-deposition of objects of different sizes, an extreme case of RSA of mixtures, with one of the components depositing much faster than another. Deposition processes of mixtures, as well as deposition on finite-size lattices and on randomly covered lattices, have been considered in the literature, e.g. [37, 24, 57–67]. In this section we outline an exact solution [35] for certain deposition processes on the Bethe lattice with the substrate pre-treated by either random blocking of a fraction of sites or by a more specific preparation that allows only for monomer and dimer landing slots.

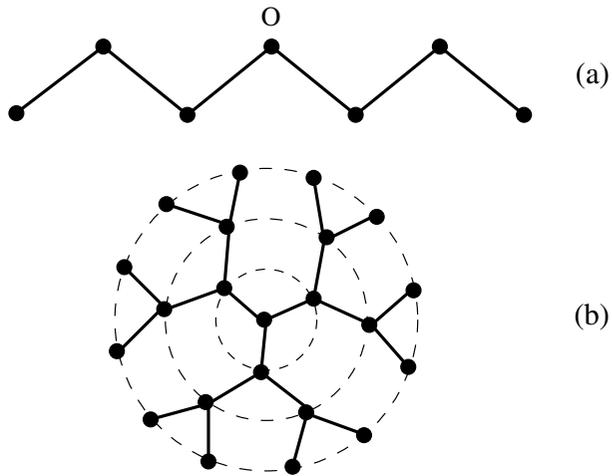

**Figure 3.** The first three 'generations' of infinite Bethe lattices originating at the central site O for (a) $z = 2$ and (b) $z = 3$.

The Bethe lattice of coordination number $z = 2, 3, \ldots$, has each site connected by bonds to $z$ nearest neighbour sites, and there are no closed loops formed by these bonds. This no-loop (no-return) property usually makes $z > 2$ Bethe-lattice results descriptive of high-dimensional behaviour. The cases $z = 2, 3$ (the former is equivalent to the 1D lattice) are illustrated in figure 3. With no loops present, one can show that each $s$-site cluster is connected by exactly $s - 1$ internal bonds and that the number of bonds shared by the $s$ cluster sites and the nearest neighbour sites immediately outside this cluster is $sz - 2(s - 1)$.

We consider RSA of dimers, arriving at the rate $\Gamma$ (per unit time) and per each nearest-neighbour pair of sites, and depositing provided both of these sites are empty. We also allow deposition of monomers, arriving at the rate $\Upsilon$ per unit time and per lattice site, and depositing only at those sites that are empty. Arriving particles that cannot be deposited are discarded. A standard approach to solving RSA problems [42] involves the consideration of the probabilities, $P_s(t)$, that a randomly chosen $s$-site connected cluster is empty. The occupancy of the neighbour sites of the cluster is not accounted for in defining this probability: some of the clusters that are empty can be parts of larger empty clusters. Note that

$$\theta(t) = 1 - P_1(t). \tag{7}$$

The Bethe-lattice property that the number, $s - 1$, of dimer landing options inside the $s$-cluster and the number, $sz - 2s + 2$, of options for landing partly inside (one site of the dimer landing at an external neighbour site) are only dependent on $s$, and not on the specific cluster configuration, allows us to write rate equations for the empty-cluster probabilities, $P_s(t)$, which in some cases are solvable:

$$-\frac{dP_s}{dt} = \Upsilon s P_s + \Gamma[(s - 1)P_s + (sz - 2s + 2)P_{s+1}]. \tag{8}$$

One way to pre-treat the substrate is by randomly blocking some sites for deposition, with the initial fraction of the remaining empty sites given by $0 \leqslant \rho \leqslant 1$. In a sense, this is equivalent to rapid initial deposition of monomers, followed by mixture deposition. The initial conditions for (8) in this case are $P_s(0) = (1 − \rho)^s$, and the set of equations can be solved to yield

$$\theta(t) = 1 - (1-\rho)e^{-\Upsilon t}\left\{1 + \frac{\Gamma}{\Gamma+\Upsilon}(z-2)(1-\rho)[1-e^{-(\Gamma+\Upsilon)t}]\right\}^{\frac{z}{2-z}}. \quad (9)$$

This expression has interesting limiting properties [35, 61, 62] as $z \to 2$, i.e. the one-dimensional case, and for $\Upsilon = 0$, the latter corresponding to the dimer-only deposition and formation of a typical RSA jammed state with the final coverage less than unity, whereas we have $\theta(\infty) = 1$ for any $\Upsilon > 0$.

Random coverage, just considered, is an important case of pre-treated substrates that finds experimental realizations [68–71]. However, to actually control the particle deposition for self-assembly, we have to consider much more restrictive situations, when the substrate is pre-treated in such a way that particle deposition is possible only in specific locations tailored to their sizes. Let us, therefore, consider the initial conditions corresponding to only voids of size $s = 1$ and 2 left uncovered.

We will denote by $\sigma$ the initial fraction of sites that are single-site voids, and by $\tau$ the initial fraction of sites that, pair-wise, form two-site voids. Then, we have $P_1(0) = \sigma + \tau$, $P_2(0) = \tau/z$, as well as $P_{s>2}(t \geq 0) \equiv 0$. The rate equations for $P_i(t)$, $i \in \{1, 2\}$, then yield an expression which, not surprisingly, is not sensitive to the coordination number $z$,

$$\theta(t) = 1 - e^{-\Upsilon t}\left\{\sigma + \tau + \frac{\Gamma}{\Gamma+\Upsilon}[1-e^{-(\Gamma+\Upsilon)t}]\right\}. \quad (10)$$

## 5. RSA on a pre-patterned 2D lattice

In this section, we outline some of our recent Monte Carlo simulation results [36] on how particles of spherical shape (circular cross-section) irreversibly adsorb on substrates patterned with a square grid of square regions onto which the particle (projected) centres can adhere. We consider certain correlation properties of the jammed state, as one varies the various length-scales involved. We specifically chose a pattern of squares, because of their experimental interest, though other shapes can also be considered [3]. Thus, the 'sticky' portion of the substrate consists of squares of size $a \times a$, with their parallel sides separated by a distance $b$. One can relate, without loss of generality, all length scales to the diameter, $d$, of the depositing circular-projection particles. Therefore, we have two effective parameters, namely,

$$\alpha = a/d \quad \text{and} \quad \beta = b/d. \quad (11)$$

With the above assumptions, the model represents a generalized version of the random sequential adsorption (RSA) or car parking model [72, 73, 24, 27, 28]. To better classify the rich variety of cases embodied in the model, we partition the 'phase diagram' into four regions according to the distance between the squares and their size. We note that arriving particles can overlap with (be rejected due to) particles centred in different cells only for $\beta < 1$; this will be classified as the interacting cell–cell adsorption (ICCA). For $\beta \geq 1$, there is no cell–cell overlap; this will be classified as non-interacting cell–cell adsorption (NICCA). Notice that in the ICCA case, and depending on the relative size of the particles as compared to the cell (the value of $\alpha$), an arriving particle could overlap (interact) with particles in cells farther than the nearest-neighbour ones. Now regarding the size of the cells, at most a single particle can deposit in each cell for $\alpha < 1/\sqrt{2}$; this will be classified as single-particle-per-cell adsorption (SPCA). For $\alpha \geq 1/\sqrt{2}$, more than one particle can be adsorbed in each cell; this will be classified as multiparticle-per-cell adsorption (MPCA). These regions in the $\alpha\beta$ plane are shown in figure 4.

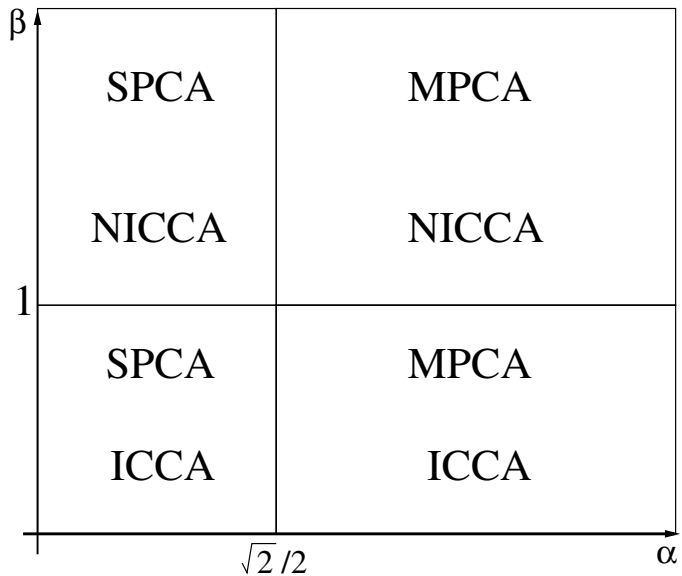

**Figure 4.** Phase diagram of the model.

Obviously, the relative values of $\alpha$ and $\beta$ determine the extent to which the square pattern affects jamming effects in particle (centre) deposition with each cell, as well as the jamming properties due to particles in neighbouring cells. A rich diagram of various possibilities can be developed to expand the classification shown in figure 4, though ultimately numerical simulations will be needed to clarify the significance of the 'phase boundaries' suggested by such geometrical considerations. For values of $\alpha \to 0$ with $\beta \geqslant 1$, the model reduces to RSA of monomers on a square lattice of spacing $b$. There are then no jamming effects at all. However, for $\alpha \to 0$ with $\beta < 1$, we obtain deposition of extended objects, since each deposited particle 'shadows' several lattice sites. For $\alpha \to \infty$ or $\beta \to 0$ (with the other parameter fixed), one recovers the usual continuum RSA model in 2D.

We outline here two particular cases, with representative values for the $\alpha$ and $\beta$. Our pattern corresponds to a square lattice of cells (with their no-deposition boundary regions) of linear size $\alpha + \beta$. For good statistics, simulations were carried out for substrates consisting of $500 \times 500$ such cells, with periodic boundary conditions applied both vertically and horizontally. We simulated $10^3$ time steps, where a time step was defined as the time required to have a number of deposition attempts, on average, that would yield a packed monolayer of particles, were they actually all deposited and repositioned to form a closed-packed structure. The Monte Carlo results were averaged over $10^2$ independent runs.

To characterize the jammed state, illustrated in figure 5, we studied the radial distribution function [36] of the distances between the particle centres, $\mathcal{P}(r, \alpha, \beta)$. The shape of the radial distribution function is significantly affected by the values of $\alpha$ and $\beta$. For $\alpha = 0.6$ and $\beta = 0.2$, the distribution of the particles is homogeneous even on length scales comparable with their size, as is seen in figure 5(a). Indeed, figure 5(b) demonstrates a nearly featureless radial distribution function. The peaks observed in the radial distribution function correspond to distances defined by the cell positions in the square lattice matrix, as marked in figure 5(c). On the other hand, for $\alpha = 0.2$ and $\beta = 0.5$, the jammed state shows local order, as seen in figure 5(d). The radial distribution function now shows a series of well developed peaks, see figure 5(e), which correspond to the cell-defined distances in the square lattice matrix, as shown in figure 5(f). This emergence of the local order is a correlation effect that develops during the deposition stage, due to the pre-patterning of the substrate.

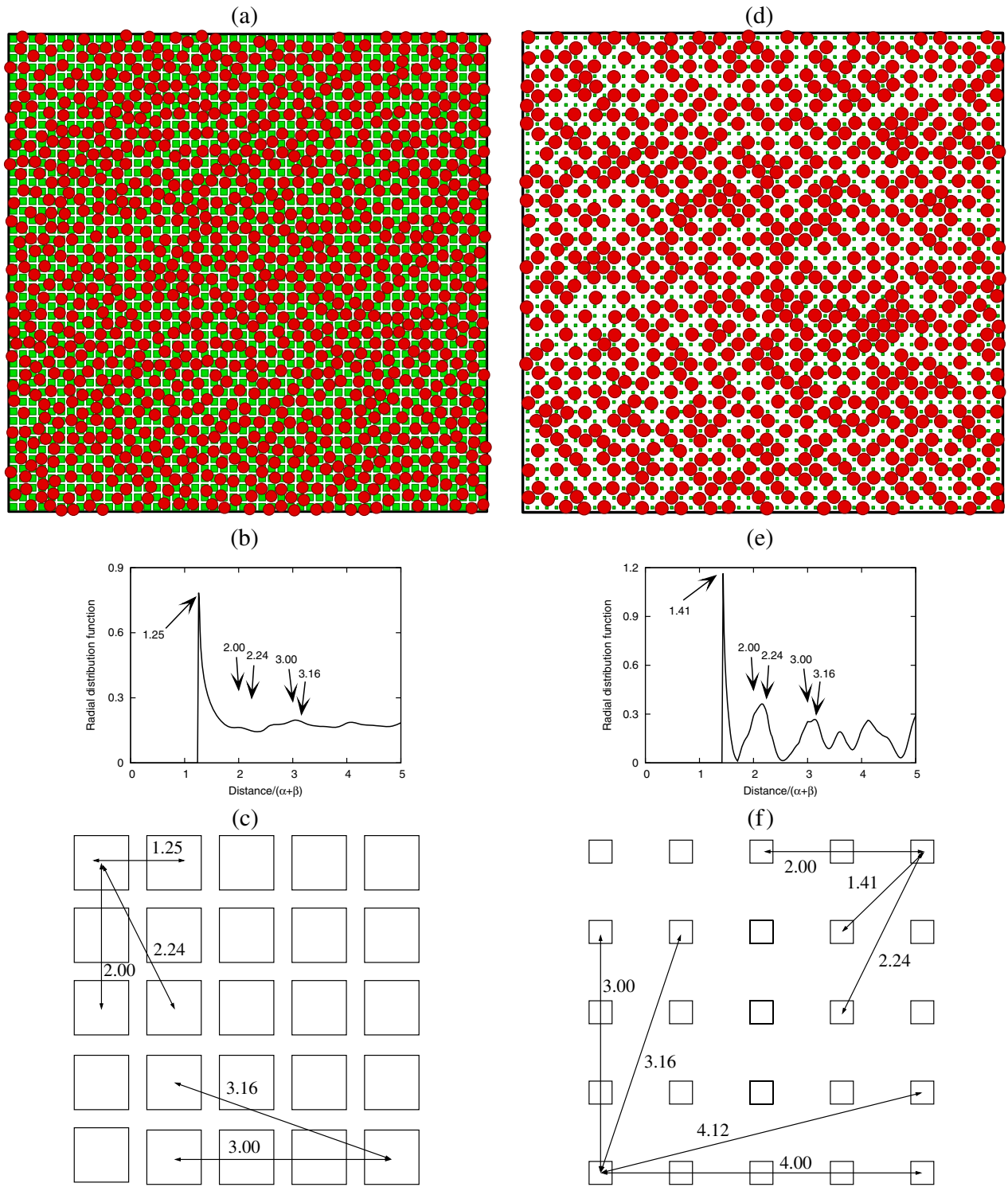

**Figure 5.** Simulation for two choices of the parameters $\alpha$ and $\beta$, for a periodic-boundary-condition system of size $500 \times 500$ pattern-cells: (a) example of a jammed state for $\alpha = 0.6$, $\beta = 0.2$, with (b) the corresponding radial distribution function, and (c) the pattern-defined distances that correlate with the peaks in the radial distribution; (d) a jammed state for $\alpha = 0.2$, $\beta = 0.5$, with the corresponding (e) radial distribution function and (f) pattern-defined distances.

(This figure is in colour only in the electronic version)

In summary, the versatility of the RSA model has been illustrated by the range of phenomena it has been useful in explaining, since the seminal work of Flory in 1D [72]. The model has been extended in several directions, most notably to include the kinetics [42, 27, 24], as well as being studied in higher spatial dimensions for both on- and off-lattice variants.

Functional forms explaining the approach to the limiting coverage depend on the particle shape in the off-lattice case. On-lattice versions show an exponential approach to the limiting coverage at late times. However, for large values of $k$ (the number of lattice distances that fit in a particle size), an intermediate regime, for $t \simeq (RA)^{-1}$, is found, with an off-lattice type kinetics. The off-lattice (continuum) regime is in fact obtained exactly by taking the limit $k \to \infty$. Versions of the continuum RSA model with deposition of mixtures of particles of different sizes and shapes have only been studied for the simplest cases of two particle sizes and identical shapes [67, 74, 75, 66, 35, 76–78]. In spite of all the research work thus far, the field remains widely open to new research efforts. Specifically, correlations have to be further studied, especially for the jammed state and on approach to it.

As experiments on submicron down to nanometre length scales are becoming more common, the need for studies involving control of particle positioning has also emerged. Among the various possibilities, patterning of the surface represents a promising option, but more theoretical work is needed. Specifically, there are no available studies of the effects of different pattern structures (from irregular to lattice), pattern shapes, and particle shapes; all these may introduce new interesting features. In the present work, we reviewed the deposition of discs on square cells arranged on in a square-lattice pattern. In one case the choice of the parameters led to a locally relatively homogeneous deposition of the particles, while in another case a degree of local order was imposed.

## Acknowledgments

This research was funded by two Fundação para a Ciência e a Tecnologia research grants: Computational Nanophysics (under contract POCTI/CTM/41574/2001) and SeARCH (Services and Advanced Research Computing with HTC/HPC Clusters) (under contract CONC-REEQ/443/2001) and by the US National Science Foundation (grant DMR-0509104). One of us (NA) also thanks Fundação para a Ciência e a Tecnologia for a PhD fellowship (SFRH/BD/17467/2004).